\documentclass[review]{elsarticle}
\usepackage[title]{appendix}
\usepackage{lineno,hyperref}
\usepackage{amstext}
\usepackage{graphicx}
\usepackage{mathtools} 
\usepackage{mathptmx} 
\usepackage{booktabs}  % For formal tables
\usepackage{tablefootnote}
\usepackage{supertabular}
\usepackage{amsmath}
\usepackage{amssymb}
\usepackage[capposition=top]{floatrow}
\usepackage{listings}
\usepackage[most]{tcolorbox}
\usepackage{inconsolata}

\newtcblisting[auto counter]{sexylisting}[2][]{sharp corners, 
    fonttitle=\bfseries, colframe=gray, listing only, 
    listing options={basicstyle=\ttfamily,language=java}, 
    title=Search \thetcbcounter: #2, #1}

\newcommand{\minus}{\scalebox{0.75}[1.0]{$-$}}

\modulolinenumbers[5]

\journal{Information \& Software Technology}

\begin{document}

\begin{frontmatter}

\title{The impact of using biased performance metrics on software defect prediction research}

\author[mymainaddress]{Jingxiu Yao}
\ead{JingxiuYao@buaa.edu.cn}

\author[mysecondaryaddress]{Martin Shepperd\corref{mycorrespondingauthor}}
\cortext[mycorrespondingauthor]{Corresponding author}
\ead{martin.shepperd@brunel.ac.uk}

\address[mymainaddress]{Beihang University, Beijing, China}
\address[mysecondaryaddress]{Brunel University London, UK}

\begin{abstract}
\emph{\textbf{Context}}: Software engineering researchers have undertaken many experiments investigating the potential of software defect prediction algorithms.  Unfortunately some widely used performance metrics are known to be problematic, most notably F1, but nevertheless F1 is widely used. \newline
\emph{\textbf{Objective}}: To investigate the potential impact of using F1 on the validity of this large body of research. \newline
\emph{\textbf{Method}}: We undertook a systematic review to locate relevant experiments and then extract all pairwise comparisons of defect prediction performance using F1 and the unbiased Matthews correlation coefficient (MCC). \newline
\emph{\textbf{Results}}: We found a total of 38 primary studies.  These contain 12,471 pairs of results.  Of these comparisons, 21.95\% changed direction when the MCC metric is used instead of the biased F1 metric.  Unfortunately, we also found evidence suggesting that F1 remains widely used in software defect prediction research.\newline
\emph{\textbf{Conclusions}}: We reiterate the concerns of statisticians that the F1 is a problematic metric outside of an information retrieval context, since we are concerned about both classes (defect-prone and not defect-prone units).  This inappropriate usage has led to a substantial number (more than one fifth) of erroneous (in terms of direction) results.  Therefore we urge researchers to (i) use an unbiased metric and (ii) publish detailed results including confusion matrices such that alternative analyses become possible.
\end{abstract}


\end{frontmatter}

%\linenumbers

\section{Introduction}\label{intro}
\noindent
The idea of trying to predict which software components e.g., classes, files or even methods are likely to be defect-prone has gained a great deal of traction in software engineering research over the past three decades.  To be able to reliably distinguish between defect-prone and clean components is clearly desirable since QA resources can then be allocated more effectively.  A considerable number of systematic reviews \cite{Cata09,Hall12,Malh15,Hoss17,Ozak18,Son19,Li20} have identified and summarised many hundreds of such studies.

These defect prediction studies have generally approached the problem empirically in the form of computational experiments, where different prediction systems are compared over data with known outcomes (i.e., labelled defect-prone or not defect-prone).  Comparisons are made using various classification performance metrics, typically F1\footnote{More correctly, F1 is a specific instantiation ($\beta=1$) of the F-measure which is defined as:
\begin{equation*}
F_\beta = (1 + \beta^2) \cdot \frac{(\text{Precision} \cdot \text{Recall} )}{ (\beta^2 \cdot \text{Precision})+ \text{Recall} }
\end{equation*}
\noindent
However, it is a near universal practice in software defect prediction to set $\beta=1$ so for simplicity in this paper, we will simply refer to F1.} and Area under the Curve (AUC) as the response variables.  Unfortunately despite its widespread use, statisticians and machine learning specialists, have drawn attention to various difficulties with F1 particularly when it is used for two-class problems \cite{Soko09,Powe11,Luqu19}.  Section \ref{Subsec:CritF1} reviews these difficulties in some detail.

So we pose the question: do the problems with F1 actually matter, or are the results from the many studies utilising F1 good enough approximations to the `truth'?  This is an important question because the use of F1 remains widespread (more than a third of papers over the period 2015-20 - see Section~\ref{Subsec:F1Used}).   To answer it, we locate studies that report results both with F1 and also the unbiased Matthews correlation coefficient (MCC) \cite{Bald00} which enables us to make comparisons and check whether the conclusion changes depending upon the choice of classification performance metric.

This paper extends our previous analysis \cite{Yao20} in five ways.
\begin{itemize}
\item We provide a comprehensive review and critique of the widely-used, classification performance metric F1 drawing from both the machine learning and statistical literature, complemented with an analysis of all permutations of the N=40 confusion matrix.
\item We have extended the searches for relevant, defect prediction studies which increases the number of primary studies from 8 to 38 meaning that the number of individual results in the meta-analysis is now 12,471 (a more than threefold increase).
\item We investigate how widely the F1 metric is used in software defect prediction experiments.
\item We undertake an in-depth investigation of the circumstances when the metric F1 is most likely to be misleading, in particular imbalance.
\item The updated raw data and code are available from \url{https://doi.org/10.5281/zenodo.4899090}.
\end{itemize}

The remainder of this paper is structured as follows.  Section~\ref{Sec:Background} first reviews different approaches to evaluating defect prediction performance, followed by a detailed critique of the F1 metric and an assessment of its role within software defect prediction research.  Section~\ref{Sec:SysRev} describes the systematic review we undertook to find as many F1 and MCC results as possible, for our meta-analysis.  Then, we present our bibliometric findings in Section~\ref{Sec:BibSummary}.  This is followed by our meta- analysis, results and discussion contained in Section~\ref{Sec:Meta}.  The article concludes (Section~\ref{Sec:Concl}) with a summary of our findings, threats to validity and a set of recommendations for researchers and for readers of defect prediction studies.

\section{Background} \label{Sec:Background}
\noindent
In this section we review the most widely utilised classification performance metrics (for more exhaustive reviews see Powers~\cite{Powe11} and Luque et al.~\cite{Luqu19}).  Next, we focus on the F1 measure in detail, highlight some problems and contrast it with an alternative metric, namely the Matthews correlation coefficient.  Finally, we briefly examine the question of how widely F1 is used in software defect prediction experiments.

\subsection{Classification performance metrics in software defect prediction} \label{Subsec:PerfMetrics}

In this discussion we focus on two-class classification problems as per all of our included studies.  This approach restricts the analysis to defect-prone (positive) and not defect-prone (negative) software components.   These classes are mutually exclusive.  Thus the structure of the classifier performance can be represented as a confusion matrix (see Table~\ref{Tab:confusion_matrix}) which, since we have two classes, is a $2 \times 2$ contingency table of predicted versus actual class.  From this matrix we are able to derive most classification performance metrics.

\begin{table}
\renewcommand\arraystretch{1}
\centering
\setlength{\tabcolsep}{3pt}
\caption{The confusion matrix}
\label{Tab:confusion_matrix}
\begin{tabular}{l|cc}
\hline
 &  Actual Positive  & Actual Negative \\
\hline
Predicted Positive&
$TP$ &  $FP$\\
\hline
Predicted Negative&
$FN$ & $TN$\\
\hline
\end{tabular}
\end{table}

The four cells of the confusion matrix comprise counts of true positives (TP), true negatives (TN), false positives (FP) and false negatives (FN).  For our problem domain these correspond to defective components correctly classified as defective, defect-free components correctly classified as defect-free, defective components incorrectly classified as defect-free and defect-free components incorrectly classified as defective.  

For the discussion regarding confusion matrices we use the following terminology and concepts.
\begin{description}
\item[\emph{Prevalence:}] $p$ is the proportion of the actual positive cases (i.e., defect density) hence the prevalence of negative cases is simply $(1-p)$.  Alternatively $p = (TP + FN)/N$ where $N$ is the cardinality.  This is an important concept since it gives rise to problems of imbalanced data sets which in turn cause difficulties for training classifiers.  For the problem domain of software defect prediction, it is common, but not invariably so, that $p$ is close to zero and the datasets are imbalanced \cite{Wang13}.  Unfortunately this also leads to difficulties with biased classification performance metrics as we will shortly demonstrate.
\item [\emph{Precision:}] is defined as $TP/(TP+FP)$ which is the proportion of correctly identified defect-prone units from all cases classified as defect-prone.
\item [\emph{Recall:}] also referred to as sensitivity or the true positive rate (TPR).  It is the proportion of positive cases that are correctly predicted positive (defect-prone) out of all positive cases.  
\item [\emph{Specificity:}] or the true negative rate (TNR) is defined as the proportion of negative cases that are correctly considered as negative from all negative cases.  Specificity and Recall are inversely proportional to each other. When we increase Specificity, Recall decreases and vice versa.
\item [\emph{False positive rate (FPR):}] is defined as the proportion of negative cases that are mistakenly considered as positive out of all negative cases.  It is also sometimes referred to as Fallout and is a way of characterising the contamination of positive predictions by negative examples.  In the situation of software defect prediction this means the proportion of wasted QA effort examining components that are in reality defect free.
\item [\emph{Accuracy:}] is defined as the proportion of cases correctly classified from all cases.  However, it is not chance-corrected and therefore often a misleading guide.  Consider the situation where 95\% of cases are positive (so $p=0.95$) then trivially a classifier could achieve 95\% accuracy simply by predicting all cases belong to the modal (in this case positive) class.
\item [\emph{F1:}] is the harmonic mean of Precision and Recall. It is based on the F-family of measures, but specifically where $\beta=1$ regulates Precision and Recall such that they have equal weight.  Although Precision and Recall can each be trivially optimised independently (either predicting no positive cases or by predicting all positive cases) the idea of combining both measures is intended to take a more balanced view.  For this reason F1 has been widely deployed as a means of assessing classifier performance.  NB A harmonic mean, as opposed to an arithmetic mean, will penalise more extreme differences between Precision and Recall.
\item [\emph{Bookmaker's odds:}] is also known as Youden's J \cite{Youd50} or Informedness for multi-class classification. It is defined as: $\text{Recall} + \text{Specificity} -1$.  It yields the proportion of time we are making an informed decision as opposed to guessing \cite{Powe03}.  A classifier that has a Bookmaker score of zero is doing no better than chance and a negative score implies worse than chance.  This is important information when evaluating classifiers.  Its chief value is as a simple benchmark of the extent to which a classifier adds value.
%\item[\emph{Brier score:}] 
\item [\emph{Matthews correlation coefficient (MCC):}] is the Pearson correlation for a contingency table and is known as $\phi$ or $r_{\phi}$ by statisticians. It is defined using TP, TN, FP and FN, and so includes all parts of the confusion matrix.  As with any correlation coefficient, it ranges  from -1 to +1 so more extreme values represent better performance.  Thus +1 indicates perfect classification, -1 indicates perfectly perverse classification, and zero indicates random predictions i.e., no classification value.  It is related to the chi-square statistic for a $2 \times 2$ contingency table such that $\phi = \sqrt{\frac{\chi^2}{n}}$ where $n$ is the number of cases.  
%\item The G-mean (GM) is calculated as the geometric mean of $TPR$ and $1-FPR$.
\item [\emph{Receiver operating characteristic (ROC) curve:}] is a two-dimensional chart where the TPR (Recall) is plotted on the Y axis and FPR (Fallout) is plotted on the X axis.  A ROC curve describes the relative trade-offs between TPs and FPs for different thresholds of accepting a case as being positive ranging from all (when TPR=1) to none (when FPR=0). As a two-dimensional assessment index, it can be problematic when comparing the performance of different classifiers. In order to compare classifiers using a single scalar value, the usual method is to calculate the area under the ROC curve (AUC) \cite{Fawc06}.  Since AUC is a proportion of the total area of the unit square, its value must fall between 0 and 1, where AUC=1 means the classifier can perfectly distinguish between positive and negative classes, whilst AUC=0 means the classifier is perfectly perverse, i.e., it predicts all positive cases as negative ones and vice versa.  When AUC=0.5, the classifier has no discriminative value and is equivalent to random guessing (equivalently Youden's J or MCC equal to zero).  Therefore, any value greater than 0.5 represents a better than chance classification for the two-class case.  Note, however, this metric refers to a family of possible classifiers rather than any specific classifier. Thus, unless the ROC curve for classifier A strictly dominates classifier B we cannot make any remarks about our preference for A over B since, in practice, we can only deploy a single classifier.  Morasca and Lavazza \cite{Mora20} suggest this difficulty might be reduced by only examining "relevant areas", that is regions of interest.   But we also note that  AUC has come under considerable criticism (for example that it uses different misclassification cost distributions for different classifiers \cite{Hand09,Flac15}).  Moreover, its purpose is somewhat different from our primary interest which is to address performance metrics for \emph{particular} classifiers.  For this reason we do not explore AUC further in this paper. 

\end{description}

\noindent
Table~\ref{Tab:perf_metrics} gives formal definitions of the most commonly deployed classification performance metrics in terms of the confusion matrix.  It also denotes those metrics that are chance-corrected, in other words making a comparison with a guessing strategy, where an example would be a negative MCC score.  By contrast, F1 is not chance-corrected because the F1 value of guessing all predictions are the modal class will depend upon its prevalence.

\begin{table}
\renewcommand\arraystretch{1}
\centering
\setlength{\tabcolsep}{3pt}
\caption{Commonly used classification performance metrics}
\label{Tab:perf_metrics}
\begin{tabular}{lcccc}
\hline
\textbf{Metric}&  
\textbf{Definition}&
\textbf{Range}&
\textbf{Better}&
\textbf{Chance corrected}\\
\hline
Cardinality, $N$ & $TP+FP+FN+TN$ & $[0, \infty]$ & n.a. & n.a.\\
\hline
Prevalence or $p$ & $p = \frac{TP + FN}{N}$ & [0,1] & n.a. & n.a.\\
\hline
Accuracy&
$\frac{TP+TN}{N}$&
[0,1]&
High& No\\
\hline
Precision&
$\frac{TP}{TP+FP}$&
[0,1]&
High&No\\
\hline
Recall or &
$\frac{TP}{TP+FN}$&
[0,1]&
High&No\\
True Positive Rate (TPR)&&&& \\
\hline
Specificity (TNR) &
$\frac{TN}{TN+FP}$&
[0,1]&
High & No\\
\hline
False Positive Rate (FPR)&
$\frac{FP}{TN+FP}$&
[0,1]&
Low&No\\
\hline
Bookmaker's odds or&
$Recall + Specificity -1$&
[-1,1]&
High & Yes\\
Youden's J & $= TPR + TNR -1$&&&\\
\hline
F-measure (F1)&
$\frac{2\times Recall\times Precision}{Recall+ Precision}  =  \frac{2 \cdot TP}{2 \cdot TP+FP+FN}$&
[0,1]&
High&No\\
\hline
Matthews correlation&
$\frac{TP\times TN- FP\times FN}{\sqrt{(TP+FP)(TP+FN)(TN+FP)(TN+FN)}}$&
[-1,1]&
High&Yes\\
coefficient (MCC) or $r_{\phi}$&&&&\\
\hline
Area Under Curve (AUC)&
FPR versus TPR &
[0,1]&
High& Yes\\
\hline
\end{tabular}
\end{table}

\subsection{A critique of F1} \label{Subsec:CritF1}
% Also mention the Boyd et al paper?
F1 is a widely used performance metric in the field of software defect prediction.  It is the harmonic mean of recall and precision (see Table~\ref{Tab:perf_metrics}) and is a specific instantiation ($\beta=1$) of the F-measure that is defined as:
\begin{equation}
F_\beta = (1 + \beta^2) \cdot \frac{(\text{Precision} \cdot \text{Recall} )}{ (\beta^2 \cdot \text{Precision})+ \text{Recall} }
\end{equation}

\noindent
It originates from the information retrieval community and was first proposed by van Rijsbergen \cite{vanR79} in the 1970s.  This metric is only sensitive to the positive class as the definition does not include TNs.  Precision is the proportion of true positives in the cases predicted positive and recall is the proportion of positive cases that are predicted.  Hence their values are entirely \emph{independent} of the number of negative cases.  Ignoring negative cases, except inasmuch as they contaminate predictions, makes perfect sense when the problem domain is essentially a \emph{single-class problem} e.g., retrieval of relevant pages from the web when the number of irrelevant pages correctly not retrieved (TNs) is vast, cannot be determined and is not of interest.  

So we have a classification metric suitable for single-class information retrieval problems being redeployed for two-class problems.  Moreover, we speculate that some researchers have not fully considered the ramifications of doing so.  For software defects (and many other problem domains) we most definitely have two classes.  Knowing that a software unit has been correctly classified as not defect-prone is important in terms of project resources and software quality.   In addition, the dichotomous view of prediction is an over-simplification since most classifiers predict class membership with a given confidence or probability.  The threshold for positive class assignment is therefore both flexible and arbitrary in that changing the acceptance threshold for positive cases can move a software unit from being predicted positive (defect-prone) to not defect-prone.  The problem of two classes is compounded by much variation in the prevalence of defect-prone cases in training data sets.  Typically these positive cases are very much in the minority, hence most data sets are highly imbalanced \cite{Sun09}.

A second problem is that F1 is difficult to interpret other than zero\footnote{Strictly speaking, even zero can be problematic because in the event TN=0 then F1 is undefined, although it is customary to record this situation as F1=0.} is the worst case and unity the best case.  Specifically the chance component of the metric is unknown, unlike a correlation coefficient (see Table \ref{Tab:perf_metrics}).  So, for example, it is hard to know what $F_1 = 0.25$ means.  Is it better than chance?  Is the classifier actually predicting or would we be better off just guessing?  In contrast, a correlation coefficient equal to 0.25 means there is a small positive effect and that the classifier is indeed doing better than chance.

Third, F1 is not chance-corrected.  An alternative way to think about the difficulties of F1 is in terms of its relationship to the so-called Bookmaker's odds (otherwise known as Youden's J \cite{Youd50} or informedness for multi-class classification problems \cite{Powe11}).  It gives the probability that the classifier is doing better than chance (see Table \ref{Tab:perf_metrics}) and is independent of the relative proportions of positive and negative (defect-prone and not defect-prone) instances.  Youden suggested that the metric ranges [0,1] since he appears not to have considered the possibility of a perverse classifier.  Unfortunately such circumstances do arise in machine learning experiments, see for example the meta-analysis in \cite{Shep14}.  Thus, the Bookmaker's odds provides a simple benchmark\footnote{Whilst we utilise the Bookmaker's odds to be a simple benchmark to evaluate the chance component of any classification performance, we consider correlation coefficients such as MCC to be more useful for overall performance evaluation.  In the case of MCC it has the added utility of being related to a chi-squared distribution.} to assess F1.

\begin{figure}[htp]
\begin{centering}
\includegraphics[width=\linewidth]{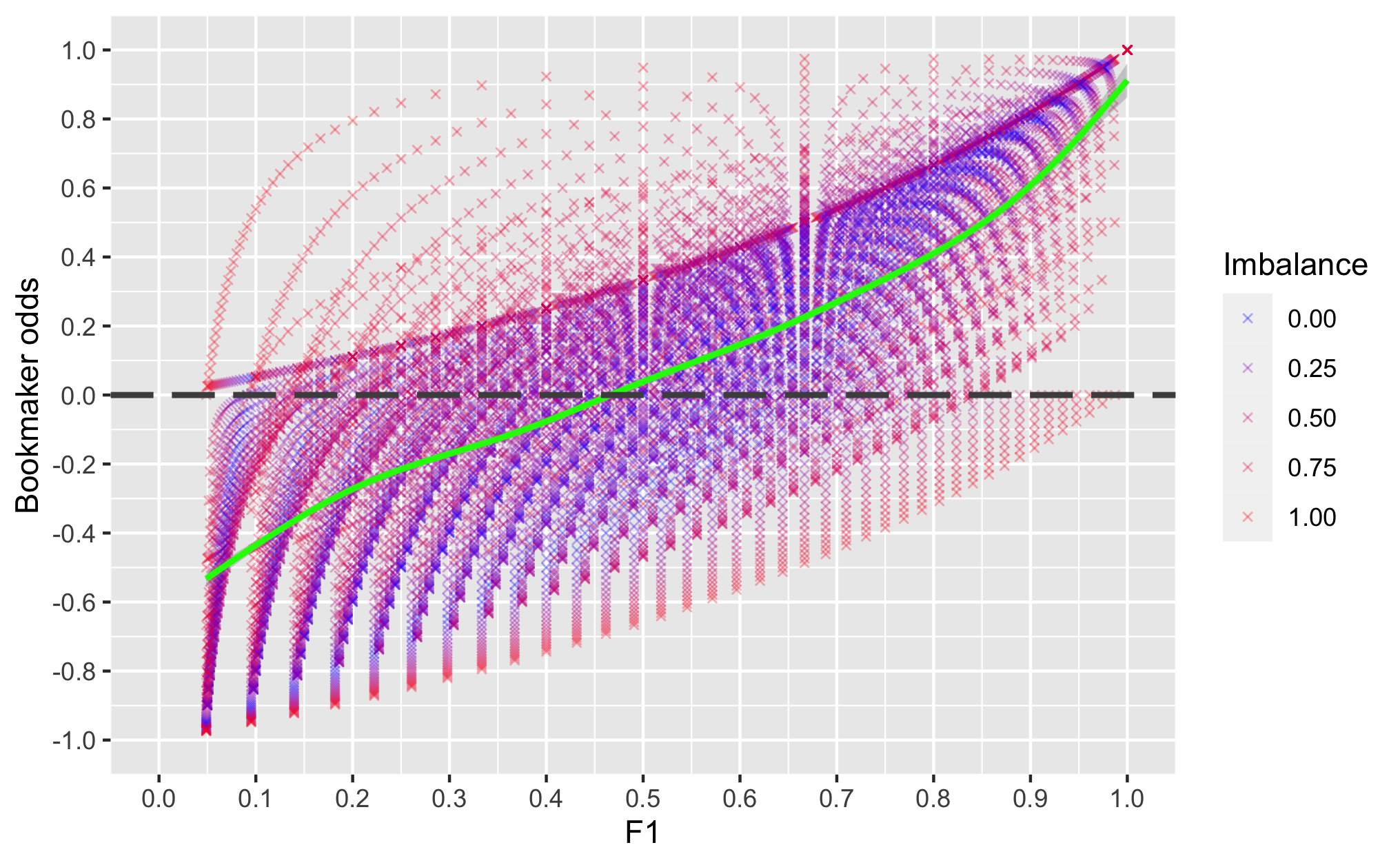}
\caption{Plot of F1 scores by Bookmaker's Odds for all valid permutations of a N=40 confusion matrix}
\floatfoot{{\scriptsize \\ Imbalance is defined as $abs(p-0.5) \times 2$ and shaded between blue (low) and red (high). The green trend line is based on a loess smoother with bootstrapped 95\% confidence limits shown as a shaded grey area.  The grey long dashed line shows Bookmaker's odds of $J=0$, below which a classifier is performing worse than chance.}}
\label{Fig:BookmakerF1}
\end{centering}
\end{figure}

In Figure~\ref{Fig:BookmakerF1} we plot, for all 12341 possible permutations of an $N=40$ confusion matrix, the F1 score, and the associated Bookmaker's odds and the degree of imbalance, defined as: $abs(p-0.5) \times 2$ where $p$ is the prevalence (of positive cases).  Note that for 940 permutations one or both metrics has no defined value e.g., because TP=0.   We can observe there is some general tendency for the Bookmaker odds to increase as the F1 score increases as indicated by the green smoothed line.  However, there are many deviations and if we examine the lower righthand quadrant (F1 $>$ 0.5 and Bookmaker $<$ 0) we see there is extensive potential for F1 to provide very misleading scores for a classifier that is in actual fact worse than random.  As an extreme example, the confusion matrix: $\left[\begin{array}{cc}35 &  1\\4 & 0\end{array}\right]$ yields F1 = 0.93 but Bookmaker's Odds of  -0.10.  In other words a near perfect F1 score corresponds to a classifier that is, in reality, slightly worse than guessing.  In other words it is perverse!  

Whilst these are hypothetical examples, it reveals a potentially, highly misleading performance metric.  We also note that these more extreme values tend to correspond to high imbalance scores (i.e., where the number of positive cases have very low or high prevalence).

Fourthly, researchers have commented on the way that F1 combines two distinct quantities, Precision and Recall and that this accomplished via the harmonic mean which will distort the impact of extreme values \cite{Hand18}.  By contrast, an arithmetic mean is more intuitively interpretable.

So we conclude that F1 is --- at least in theory --- an unreliable indicator of software defect prediction performance.  As an alternative, we propose the Matthews correlation coefficient (MCC) \cite{Bald00}, since it is chance corrected (see Table~\ref{Tab:perf_metrics}), is based upon both classes (it utilises the complete confusion matrix) and has a straightforward interpretation.  It also follows a chi-square distribution.

\subsection{How widely is F1 used in software defect prediction experiments?} \label{Subsec:F1Used}

Two previous systematic reviews have explicitly tried to quantify the extent to which F1 is used as the response variable for software defect prediction studies.  Malhotra et al.~\cite{Malh15} reported that 17/64 (~27\%) of included studies between 1991 and 2013, used F1 directly and an additional ~37\% and ~66\% used Precision and Recall, which are the two constituent components of F1 (see Table~\ref{Tab:confusion_matrix}). Another systematic review by Hosseini et al.~\cite{Hoss17} reports that out of 30 studies (2006-2016) 11 (~37\% ) use F1and 21 (~70\%) use precision and recall.

To obtain a more up to date, though somewhat approximate, view of current utilisation rates we applied the following two searches using Google Scholar (29th May, 2020).  We excluded patents and citations.

\begin{sexylisting}{General search}
"software defect prediction" 
AND 
("experiment" OR "empirical")
\end{sexylisting}

\begin{sexylisting}{F1 subset search}
"software defect prediction" 
AND 
("experiment" OR "empirical")
AND 
("F1" OR "F-measure" OR "F-score")
\end{sexylisting}

\noindent
These searches retrieved 	2250 (General search) and 978 (F1 subset search) results respectively, which suggests that in the past five years \footnote{We chose a period of five years in order to focus on recent trends, however our meta-analysis is actually based on papers from any time period.}of the order of ~43\% (978/2250) articles that discussed software defect prediction experiments also mentioned F1.  We then randomly sampled\footnote{The sampling was conducted by randomly sampling 3 papers from each page (of 10 papers) returned by Google Scholar, 97 pages in total. Overall we examined $97 \times 3=291$ papers.} 30\% of these 978 papers and read them carefully to determine if they actually employed F1 in their analysis.   We found that 82\% (239/291) of  papers actually used the F1 metric in their methods.  From this we argue that in the last five years there are of the order of $0.82 \times 978 \approx 800$ software defect prediction studies that make use of the F1 classification performance metric.  Given our concerns regarding this metric we find this somewhat worrying.

\section{The systematic review} \label{Sec:SysRev}

Next we seek to find software defect prediction studies that publish results using F1 and another more reliable metric, MCC, so that we can make comparisons.

This review was carried out in January 2021.  The goal was to locate primary studies that undertook experiments to assess software defect prediction methods on historical data sets.  Specifically, we needed papers that reported results with \emph{both} the widely used, but problematic, F1 metric and the Matthews correlation coefficient (MCC).  This would enable us to determine whether differences between these metrics are merely a theoretical concern or have real-world impact.

We conducted a basic search in our earlier conference paper \cite{Yao20} and located 8 studies.  However, for this work we decided to conduct a more in-depth search which resulted in a further 30 studies making a total of 38 papers.  The details are given below (and summarised in Table~\ref{Tab:SysRevSummary}).

\begin{table}[htp]
\caption{Systematic review summary}
\label{Tab:SysRevSummary}
\begin{tabular}{p{3cm}|p{9cm}}
\textbf{Characteristic} & \textbf{Description} \\
\hline
Objective & To find experimental results where both the F1 and MCC classifier performance metrics are reported\\
Target domain & Software defect prediction experiments \\
Target audience & (i) Ourselves (for a meta-analysis) and (ii) other researchers \\
Date & January 2021 \\ 
Databases searched & Google Scholar \\
Additional searches & Forward and backward chaining, results from previous search \cite{Yao20} \\
Inclusion validation & JY and MS independently checked papers for potential inclusion and disagreements were discussed\\
Grey literature & Not included \\
Study quality & (i) Refereed and (ii) for predictive studies uses some cross-validation procedure\\
Data collected & (i) Bibliographic data including: authors, title, year and publication venue and types of classification performance metric collected, (ii) result count, (iii) individual F1 and MCC results for $D \bigtimes C$, where $D$ is dataset and $C$ is classifier, (iv) inference procedure details including: use of NHST and correction procedures e.g., Bonferroni\\
Data published & zenodo \url{10.5281/zenodo.3949897} \\
\hline
\end{tabular}
\end{table}%

We decided to restrict our search to the domain of software defect prediction experiments.  This was because there are aspects to the researcher's choice of classifier performance metric that are domain-specific, namely (i) whether true negatives can be enumerated, i.e., is this count knowable and (ii) the prevalence of true cases, i.e., defect-prone software components.  In the case of software defects, not only is the number of true negatives knowable, it is important since these are the software components that are correctly identified as not being defect-prone.  Such components can then be allocated reduced testing resources.  The other aspect is that in general, the majority of data sets have few positive (defect-prone) cases, i.e., they are imbalanced.  These two conditions render F1 an unsafe choice of performance metric for software defect prediction experiments (see Section~\ref{Subsec:CritF1}).

We wanted to ensure that we collected high quality experimental results for our meta-analysis.  For this reason we only used peer reviewed studies, meaning that the so-called grey literature has been excluded.  Although some researchers are strongly advocating `multi-vocal' literature reviews e.g., \cite{Garo19} our purpose is slightly different.  First, we are not motivated by ``closing the gap between academic research and professional practice" (Elmore \cite{Elmo91} quoted in Garousi et al.~\cite{Garo19}) because our target is researchers.  Second, we are aware of the ease with which it is possible to make errors in computational experiments and therefore are motivated to maximise independent scrutiny. Of course, the peer reviewed literature is still replete with mistakes \cite{Dono09,Alli16} and more specifically in defect prediction \cite{Shep19,Li20}.

The complete list of exclusion criteria is given in Table~\ref{Tab:InclCrit}.  The long list of 194 articles was constructed by scanning the title and venue.  Where there was any doubt, the article remained in the long list.  If we were unable to obtain content through the usual channels we attempted to email the authors.  Each criterion was successively applied in the order listed in Table~\ref{Tab:InclCrit}, consequently the counts only refer to the residual articles (e.g., if an article is unavailable we make no judgement as to whether it is written in English).  Note `new data' refers to the situation where the same experimental results are presented in more than one paper.  Finally, `suitable data' is something of a catch all where we are unable to use the data for a range of particular reasons such as no meaningful differences between the classifiers being compared or no comparisons (i.e., results being presented without benchmarks).

\begin{table}[htp]
\renewcommand\arraystretch{0.8}
\caption{Systematic review exclusion criteria}
\label{Tab:InclCrit}
\begin{tabular}{lrr}
\hline
\textbf{Exclusion criterion} & \textbf{Removed} & \textbf{Remaining}\\
\hline
Long list of candidate papers & - & 194 \\
Content unavailable & 9 & 185\\
Not written in English & 6 &  179 \\
Different problem domain to software defect prediction & 45 & 135\\
Not refereed & 24 & 113 \\
Provides F1 \emph{and} MCC data & 69 & 43 \\
New data &  1 & 42 \\
Suitable data & 4 & 38 \\
No cross-validation (where appropriate) & 0 & 38 \\
\hline
\end{tabular}
\end{table}%

\section{Bibliometric summary data}\label{Sec:BibSummary}

As indicated, we located 38 research papers that described computational experiments that compare the performance of different classifiers, e.g., logistic regression of random forest algorithms, across various software defect data sets, e.g., Nasa MDP and Eclipse.   The publication venues are quite widely distributed, though we note there are three papers from Promise conferences, EMSE and JSS, two from IST, TSE and MECS (the Intl.\ J.\ of Modern Education and Computer Science).  In total there are 12 conference and 26 journal papers.   A complete list is given in Appendix~\ref{App:Details}.

\begin{figure}[htp]
\begin{centering}
\includegraphics[width=\linewidth]{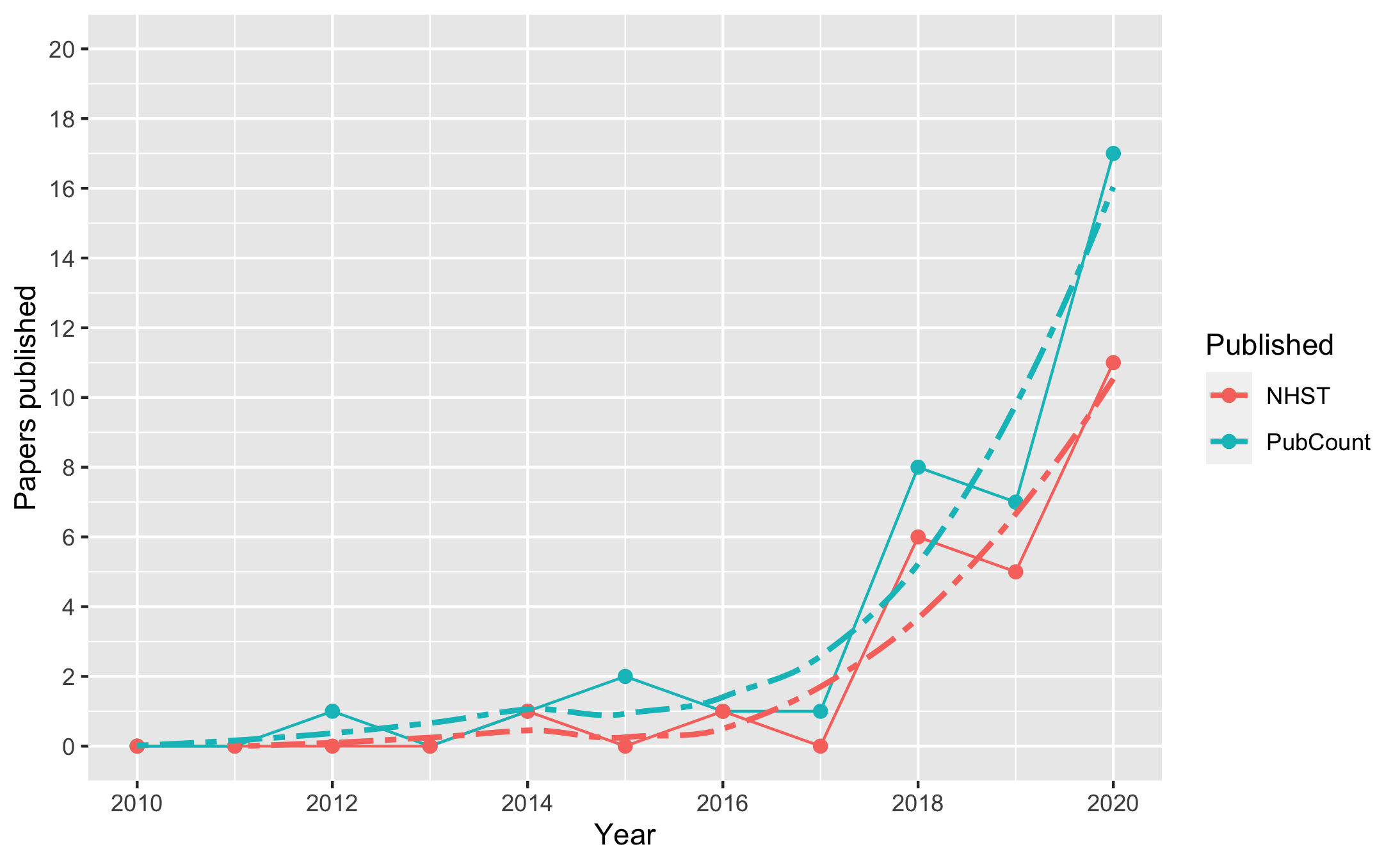}
\caption{Line plots by of papers published by year (i) reporting F1 and MCC results and (ii) making NHST-based inferences}
\floatfoot{{\scriptsize \\ NB We show counts of papers since 2010, however, our search was unconstrained date-wise.  The NHST counts refer to papers that make use of null hypothesis significance testing for making inferences about classifier comparisons.  The dashed lines are loess smoothers.}}
\label{Fig:PubTrends}
\end{centering}
\end{figure}

The papers range from 2012 to 2020.  On the whole, however, providing MCC results seems to be a relatively recent phenomenon with 32/38 papers published since 2018.  We also observed that just over half (20/38) of the papers also reported AUC, however, we did not analyse these data as being beyond the scope of our study.  \footnote{Area under the curve (AUC) would also appear to be growing in popularity amongst software defect prediction researchers.  However, it is a metric based on the frontier between the true and false positive rates and thus is a characteristic of a \emph{family} of classifiers rather than any specific classifier \cite{Fawc06,Hand09,Powe11}.}  The line plot in Fig.~\ref{Fig:PubTrends} gives an indication of the clear, steep upwards trend using a loess smoother.

In total, the 38 papers contain 12,471 usable results, that is where we have complete cases, and pairs of values for each of F1 and MCC.  The number of results per paper varied hugely from 6 to 1890 with a median of 140.5.  We noted that the papers containing a large number of results tended to be major benchmarking exercises, whilst those with few results tended to promote new algorithms or approaches and just compared these with some baseline approach.

These papers between them utilise 97 distinct datasets, however many are used multiple times across multiple studies.  Unfortunately, it would seem that different studies refer to the same data set by different names, e.g., Promise and NASA MDP.  Also, it is not always clear which version of a data set is being used plus different studies deploy differing data-cleaning strategies.  Nevertheless this indicates the breadth of research activity.

\begin{table}[htp]
\renewcommand\arraystretch{0.8}
\caption{Result decision procedure }
\begin{center}
\begin{tabular}{c|c}
Procedure & Count \\
\hline
NHST & 24 \\
Simple comparison & 10 \\
Other & 4 \\ 
\hline
Total & 38 \\
\end{tabular}
\end{center}
\label{Tab:Proc}
\end{table}%

In terms of analysis, we observe from Table~\ref{Tab:Proc} that 24/38 papers use null hypothesis significance testing (NHST) to determine whether a result is ``statistically significant" or `meaningful'.  Fig.~\ref{Fig:PubTrends} shows a clear upward trend such that in the past couple of years this is the dominant means of reasoning about comparative classification performance.  Of these, half (12/24) use some procedure (e.g., a post hoc Nemenyi test) to adjust the acceptance threshold $\alpha$ when making multiple, inferential tests. A further 10 papers merely compare values, so a higher score of the performance metric (F1 or MCC) is to be preferred to a lower score, irrespective of the magnitude.  The remaining 4 papers either use alternative methods, are unclear or have some alternative purpose than comparing competing defect predictors.

\section{Meta-analysis} \label{Sec:Meta}

\subsection{Classification performance metrics}

First we, examine the spread of values for F1 and MCC, recalling that $0 \le F_1 \le 1$ but $ \minus1 \le \text{MCC} \le 1$ so the summary statistics in Table~\ref{Tab:SummaryMetrics} are not directly comparable.  Although the statistics are derived from \emph{all} reported results from our systematic review, it is noteworthy that many classifiers appear to perform poorly.  Some even have negative correlation coefficient values (592/25,467  $\approx 2.3\%$ observations).  

As, discussed in Section~\ref{Subsec:CritF1}, the F1 metric is not easy to interpret with respect to chance odds, but we did find 235 instances of the boundary case of $F_1 = 0$ which comprise  $\approx 0.9\%$ of all observations. We assume the authors are reporting a divide-by-zero error as zero.  This could be caused by either no positive cases (TP+ FN) or no true positives (TP).  The former situation might arise from the vagaries of cross-validation (if there are few positive cases relative to the size of each fold).  The latter might arise from a very poorly performing classifier.  

\begin{table}[ht]
\renewcommand\arraystretch{0.8}
\centering
\caption{Summary statistics for all reported F1 and MCC results}
\begin{tabular}{rrr}
 \hline
 &       F1 &      MCC \\ 
  \hline
Minimum & 0.000   & -0.161   \\ 
1st Quartile & 0.296   & 0.178   \\ 
 Median &  0.400   &  0.260   \\ 
 Mean &    0.437   &  0.288   \\ 
 3rd Quartile & 0.534   & 0.360   \\ 
 Maximum &  0.999   & 0.976   \\ 
\hline
\end{tabular}
\label{Tab:SummaryMetrics}
\end{table}

\begin{figure}[htp]
\begin{centering}
\includegraphics[width=\linewidth]{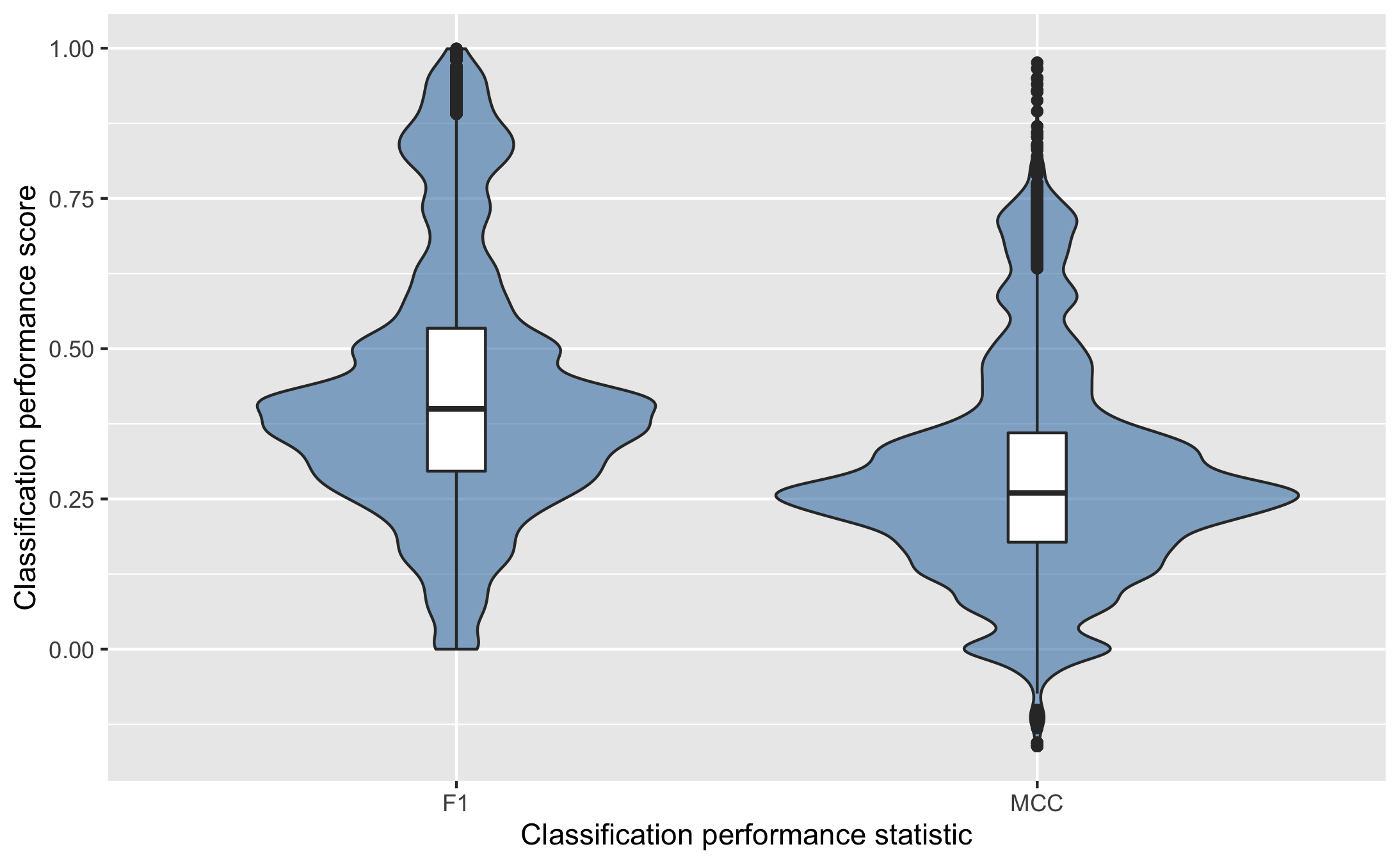}
\caption{Violin plots of the classification performance metrics F1 and MCC}
\floatfoot{{\scriptsize \\ NB the horizontal bar within the box represents the median, the box represents the interquartile range and the black dots are outliers.  Also recall, that the two metrics are measured on different scales, F1 is $\left[0,1\right]$ and MCC is $\left[{-1},1\right] $. }}
\label{Fig:Violin}
\end{centering}
\end{figure}

We can also visualise the individual distributions as violin plots (see Fig.~\ref{Fig:Violin}).  Note that, unlike MCC, the density plot for F1 is truncated at zero since this is the minimum possible value.  The kernel density estimators suggest quite complex, but positively skewed distributions, i.e., the means are greater than the medians.  In other words, most classifiers do not predict well.  However, we do stress that all results are included encompassing possibly quite na\"{\i}ve or simplistic baselines that the researchers only intended as comparators.

\begin{figure}[htp]
\begin{centering}
\includegraphics[width=\linewidth]{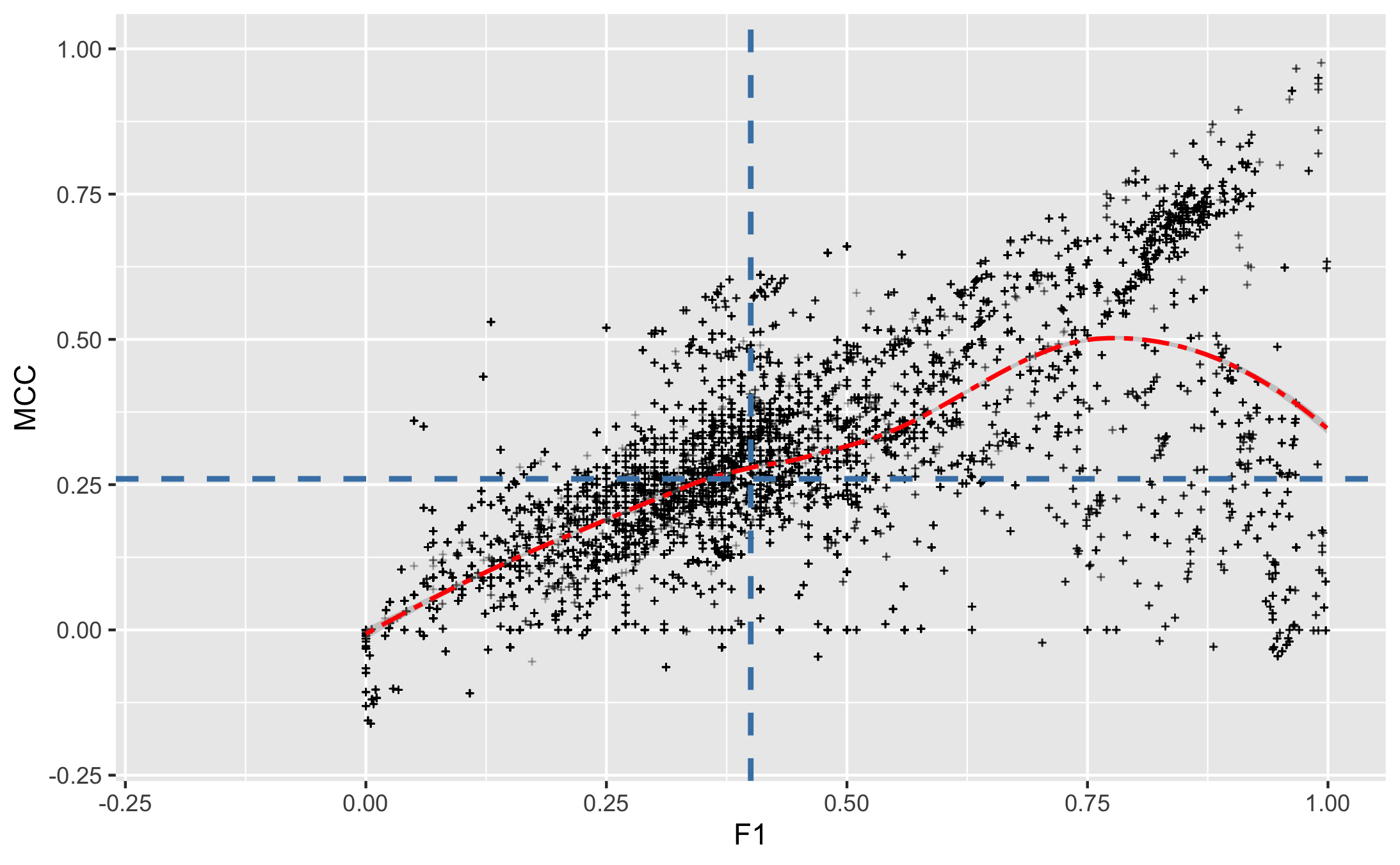}
\caption{Scatter plot of the classification performance metrics F1 vs MCC}
\floatfoot{{\scriptsize \\ NB the red-dashed line represents a loess smoother.  The blue-dashed vertical and horizontal lines represent the median values for F1 and MCC respectively.}}
\label{Fig:CorrScatPlot}
\end{centering}
\end{figure}

Although the two performance metrics are measured on different scales, we can use Spearman's correlation coefficient to evaluate the strength of a monotonic association and find $r_S=0.617$.  Some researchers would refer to this as `moderate' strength \cite{Scho18}.  Probably more informative is to examine the relationship graphically in the scatter plot given by Fig.~\ref{Fig:CorrScatPlot}.  Although the relationship is positive, i.e., as one metric value increases so does the other, we observe a good deal of scatter and some extreme outliers.  This breakdown in the relationship is most noticeable for the higher values of F1.  In other words, the relationship between the two classification performance metrics is far from straightforward and, \emph{paradoxically}, the nearer F1 is to unity the less trustworthy is the result.  A near perfect score for F1 can in practice mean anything from a similarly near perfect MCC score to a negative correlation and everything in between!  Recall, these are real, reported results in the refereed scientific literature.

\subsection{Difference between F1 and MCC} \label{Subsec:DiffF1MCC}

In order to extract information from what is a diverse set of primary studies, we must adopt a standardised view.  Essentially, all our studies are conducting computational experiments where the:
\begin{itemize}
\item \emph{treatments} are the different predictive algorithms or classifiers that are being investigated and manipulated by the researcher;
\item \emph{response variables} are the different classification performance metrics, in our case F1 and MCC;
\item \emph{experimental units} are software project data for which predictions are made;
\item \emph{block} an aggregation of similar experimental units such as a data set which is typically the level of reporting.
\end{itemize}

\noindent
Almost without exception, a repeated measures type of design is deployed by our set of primary studies, that is all treatments are applied to all experimental units.  This is possible --- unlike for many experiments involving humans --- since there is no potential for carryover or other ordering effects. Results are presented as tables of response variables, for example, see Table~\ref{Tab:ExampleReporting}. 

\begin{table}[ht]
\centering
\renewcommand\arraystretch{0.8}

\caption{A hypothetical example of reported results}
\begin{tabular}{l|rr|rr}
  & \multicolumn{2}{c|}{Logistic } &  \multicolumn{2}{c}{Na\"{\i}ve} \\
  & \multicolumn{2}{c|}{regression } &  \multicolumn{2}{c}{Bayes} \\
    Block &  F1 & MCC &  F1 & MCC \\ 
\hline
& & & & \\
Dataset1 &  0.5  & 0.4   &  0.7  & 0.6   \\ 
& & & & \\
Dataset2 &  0.3  & 0.2 &   0.8  & 0.1 \\ 
& & & & \\
\end{tabular}
\label{Tab:ExampleReporting}
\end{table}

When researchers assess their results they do so by making comparisons between treatments.  Which one is to be preferred?  These are generally decomposed to pairwise comparisons.\footnote{Although omnibus tests (e.g., a Friedman test) are sometimes utilised to make comparisons of multiple results such tests can always be decomposed to their primitive components.}   In Table \ref{Tab:ExampleReporting} we  see some hypothetical results.  In this simple example we have two comparisons, one for each block or dataset, of Logistic Regression (LR) compared to Na\"{\i}ve Bayes (NB).  For example, using F1 we have Na\"{\i}ve Bayes out-performing Logistic Regression for both data sets since ($0.7>0.5$ and $0.8 > 0.3$), however, the results using MCC are not fully concordant since ($0.6 > 0.4$ \emph{but} $0.1 < 0.2$).  In the latter case, which treatment or type of defect predictor we prefer depends upon the choice of the measurement function $m$. In such a situation we refer to the results as being \emph{discordant}.  One can imagine the hypothetical example being extended with a third classifier, say Random Forest, which would then mean the table of results could be decomposed to six comparisons (LR-NB, LR-RF, NB-RF twice over, once per dataset).  To generalise, a table reporting $t$ treatments over $d$ data sets can be decomposed to $d((t(t-1))/2)$ comparisons.

More formally, the comparisons are made using different metrics or measurement functions $m$.  For our analysis, we have $m_{\text{F1}}$ and $m_{\text{MCC}}$. These functions are applied to different dataset-treatment combinations, yielding for example, $m_{\text{F1}}$(Dataset1, LR) which gives the F1 metric from applying Logistic Regression to Dataset1.  By taking pairs of measures, we can establish preference relations, e.g.,  $m_{\text{F1}}$(Dataset1, LR) $\prec m_{\text{F1}}$(Dataset1, NB), in other words, NB is to be preferred to LR for Dataset1.  The question arises whether other measurement functions, MCC in our analysis, yield concordant or discordant relations. If the choice of metric governs the outcome this is concerning, the more so because we know that F1 is a flawed metric in the context of two-class classification in software defect prediction.

Comparisons between pairs of treatment-blocks can differ in either magnitude or direction.  Discordance, our focus, addresses the latter.  In other words using F1 leads one to prefer X to Y, yet using MCC would lead the investigator to conclude the opposite.  Researchers adopt a range of approaches when considering how evaluate differences in magnitude, particularly small differences.  A common, though controversial, approach is the use of null hypothesis significance testing (NHST) \cite{Cohe94,Gelm06,Colq14}. Here the idea is to distinguish between small differences in magnitude that might be merely due to noise, and differences that are `true'.  In such circumstances one can deploy a ``not worse than" ($\nprec$) preference relation.  Scott-Knott is another approach with a similar goal \cite{Mitt15}.

\begin{figure*}[htp]
\begin{center}
\includegraphics[width=\linewidth]{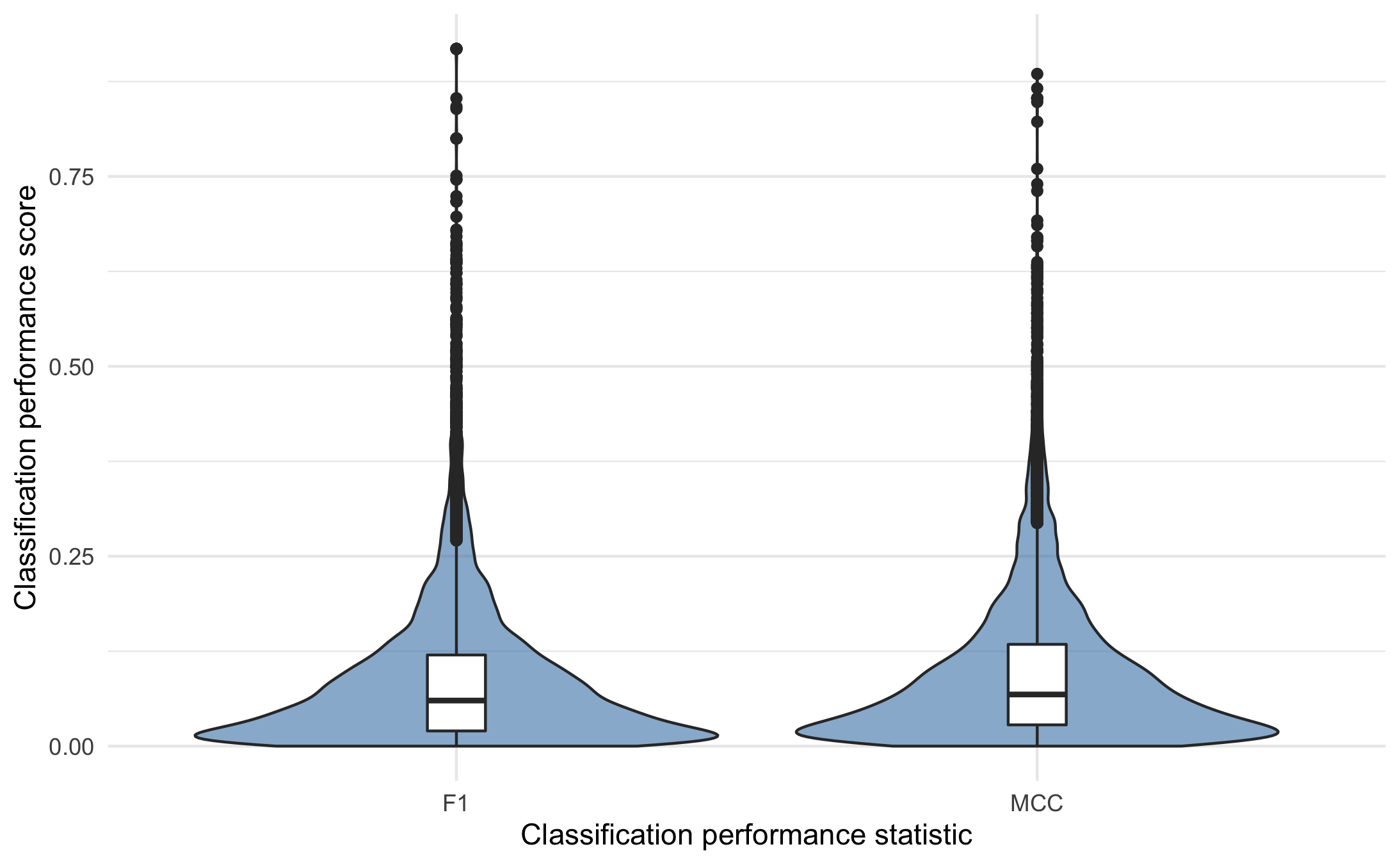}
\caption{Violin plots of the absolute differences between treatments captured by F1 and MCC}
\label{Fig:ViolAbsMetrics}
\end{center}
\end{figure*}

We are agnostic about the direction of the difference, since whether a classifier is interpreted as treatment 1 or 2 is entirely arbitrary, thus we look at absolute differences.  Figure~\ref{Fig:ViolAbsMetrics} shows the distribution of differences together with a kernel density estimator.  Both metrics show a similar highly positively skewed distribution (partly as a consequence of taking the absolute difference), nevertheless it is noteworthy that for both metrics the median differences between pairs of observations are small (F1 = 0.060 and MCC = 0.068).  This suggests that for the majority of comparisons between pairs of defect predictors, the disimilarities are small.

The question arises, if a difference in direction occurs, could this be due to trivial change in an accuracy metric (e.g., $ \langle 0.5,0.501 \rangle \text{compared with} \langle 0.501, 0.5 \rangle$)?  In other words what if many differences in pairs of observations are very small?  However, given that the majority of studies (24/38) use NHST as a decision procedure, even small differences are likely to be interpreted as meaningful due to the generally quite large data sets. \footnote{Using a simulation based on the NASA MDP data sets, which are the most widely used, a median dataset size of $n=911$, and a median standard deviation of sd = 0.03465 reported by Tran et al.~\cite{Tran19} we determined that a Welsh test on this data could detect (i.e., find statistically significant) a difference in treatments of  $< 0.01$ in F1 metric values (95\% CI  0.0065, 0.0127 with p-value = 2.204e-09).  This means that even very small differences in pairs of classification performance would be viewed as `significant' when viewed through the lens of NHST.  For our purposes this means even small differences between F1 and MCC might lead researchers to very different conclusions.}  Furthermore, another 10 papers simply make direct comparisons.  Being conservative we could argue a difference in F1 of 0.01 or more could be identified as `meaningful'.  If smaller differences were discarded this would eliminate 426 results or 426/2737 which is ~15\% of the conclusion changes, in other words, only a small proportion of direction changes that we identify, might be considered to be too small to be as conclusion changes by the researchers.  We return to this point in the threats to validity (see Section~\ref{SubSec:Threats}).

Next, we turn to the question of what proportion of reported results exhibit discordance.  This addresses the problem of how much does it practically matter that many research papers have used a biased classification metric as the response variable for their experiments.  Since the two metrics use different scales, we initially focus on direction rather than magnitude.  Overall from the 12,471 results there are 2737 (or 21.95\%) instances of a conclusion (or direction) change.  Using the Agresti-Coull \cite{Brow01} method to estimate the confidence interval of the binomial proportion, we have the following 95\% confidence interval:
\begin{equation*}
[ 0.219, 0.220], p \approx 0.2195
\end{equation*}
\noindent
Note that given the large number of observations the confidence interval (CI) is tightly defined. Therefore these results indicate that more than a fifth of the reported results contain conclusions that will change if the biased F1 metric is replaced by a less problematic metric such as MCC.

\begin{figure*}[htp]
\begin{centering}
\includegraphics[width=\linewidth]{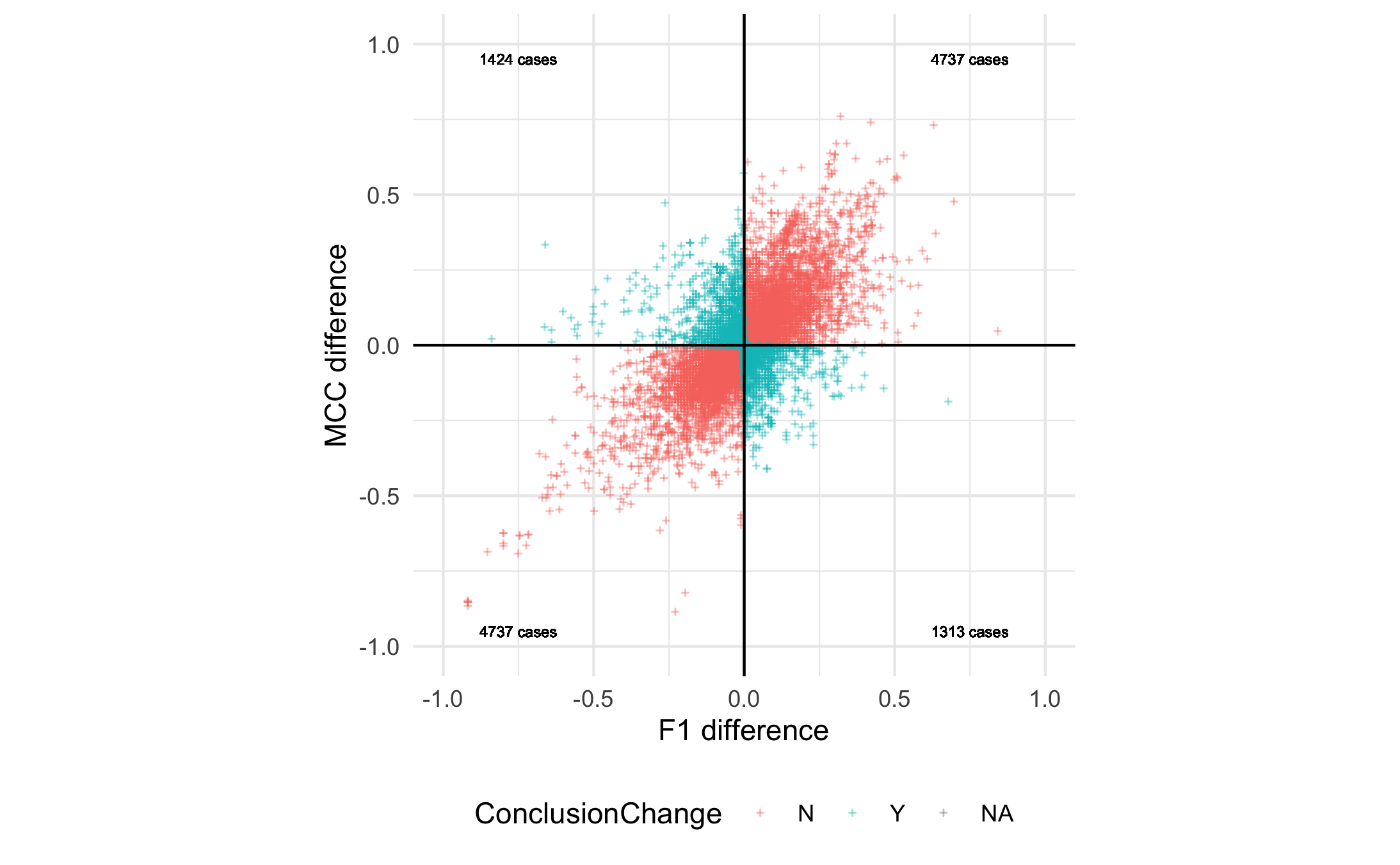}
\caption{Scatter plot of \emph{differences} between pairs of treatment effects as measured by F1 and MCC}
\floatfoot{{\scriptsize \\ The quadrants indicate where the results are concordant (coloured red) or discordant (coloured blue, thereby indicating a conclusion change since the effect direction reverses when an unbiased metric is deployed).}}
\label{Fig:QuadPlot}
\end{centering}
\end{figure*}

We can also examine the relationship between the magnitude of \emph{differences} between pairs of treatments as captured by F1 and by the unbiased MCC.  Fig.~\ref{Fig:QuadPlot} shows the comparisons for F1 and MCC plotted against each other.  When the metrics are concordant the data points fall in the lower left and upper right quadrants (coloured red) whilst a change in direction is signified when the points lie  in the upper left and lower right quadrants (coloured blue). The number of instances is given in each quadrant.  The plot shows that whilst there is a broad trend --- that as differences as captured by one metric increase so  they do for the other --- there are, however, many extreme outliers

%\subsection{Proportions of conclusion change by paper}

\begin{figure*}[htp]
\begin{center}
\includegraphics[width=\linewidth]{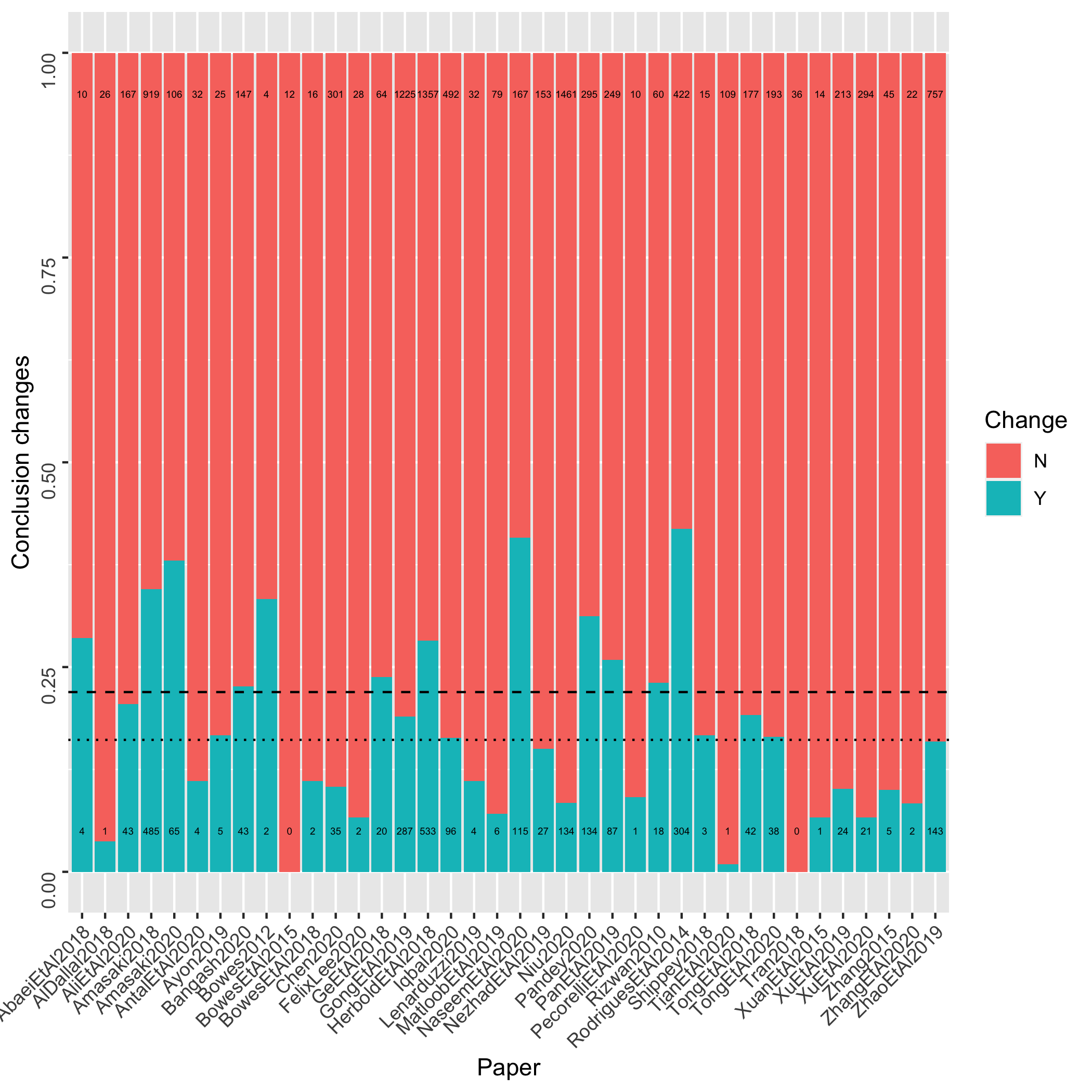}
\caption{Stacked bar plot of the proportion of results that change by paper}
\floatfoot{{\scriptsize \\ The dotted line shows the median proportion of changed conclusions by paper, whilst the dashed line shows the mean proportion of changed conclusions (21.95\%) calculated from all results.  The difference is in part due to a few papers with many results and many conclusion changes.}}
\label{Fig:sbp}
\end{center}
\end{figure*}

\subsection{Can we predict the likelihood of a conclusion change?}

Here we explore whether magnitude of the effect is an explanatory factor for the probability of a conclusion change when we replace the biased F1 classification performance metric with MCC.

%\subsubsection{Magnitude of the effect?}

One seemingly reasonable hypothesis is that the larger the observed difference between the two treatments (i.e., the effect magnitude) the less likely it matters which performance metric we employ.  In other words, a large effect size can be trusted however measured?  To investigate we use a simple procedure popularised by Gelman and Park \cite{Gelm09} based on a tertile split and comparing the bottom and top tertiles, in our case based upon absolute F1 difference (since we only care about magnitude, not direction).  The reasoning to ignore the middle tertile is to avoid comparison of adjacent items which is the well known disadvantage of a median split.
 
\begin{table}[htp]
\renewcommand\arraystretch{0.8}
\caption{Tertile analysis of conclusion change}
\begin{center}
\begin{tabular}{c|rrr}
 & \multicolumn{2}{c} {Conclusion Change} & \\
Tertile & N & Y & Odds \\
\hline
T1 & 2716 & 1364 &0.502 \\
T2  & 3495 & 943 & 0.270 \\
T3 & 3523 & 430 & 0.122 \\
\end{tabular}
\end{center}
\label{Tab:tertile}
\end{table}%
      
Table \ref{Tab:tertile} reveals a considerable difference between tertile T1 and T3 which is reflected in the odds ratio of OR = 4.11 and the 95\% CI = $[ 3.65, 4.64]$.  Again we use the Agresti-Coull method \cite{Brow01}.  From this we can see that a small effect or magnitude (in our case, an absolute difference between F1 values of $<0.031$, is as likely as not ($p \approx 0.5$) to be in the \emph{wrong} direction.  NB The effect is the absolute difference in F1 scores, as opposed to a standardised effect size such as the d-family of statistics like the widely used Cohen's $d$.  In other words, irrespective of ``statistical significance" or other arguments, a result showing a small difference in competing classifiers, as captured by F1, is as likely as not, to be in the wrong direction.

%\subsubsection{Relationship with imbalanced data set?}

\section{Conclusions} \label{Sec:Concl}

This paper has posed the question: to what extent can we rely on research results from software defect prediction studies that are based on the problematic F1 performance metric?  Unfortunately, although we, and many others before us, have shown that F1 is not a good choice of metric in the context of defect prediction (see Section~\ref{Subsec:CritF1}, F1 has been very widely used.  So, should we be concerned or is this just a minor academic quibble?  This question is the theme of our paper.

To summarise our investigation, we have searched the literature for software defect prediction studies that report performance \emph{both} as F1 and MCC.  We retrieved 38 refereed papers that contain a combined 12,471 pairs of results.  We then analysed these results by assessing whether the F1 and MCC results are concordant.  So, for example, we say a result is concordant if we prefer predictor A to B irrespective of whether we use F1 or MCC to make that comparison.   Contrariwise, we say that the results are discordant if the direction of the preference depends upon which classification performance metric is used.  That is the conclusion would change.  

Our main findings are:
\begin{enumerate}
\item Although not a new finding we show that F1 is problematic when used in a two-class problem domain such as software defect prediction.  By enumerating and then plotting all N=40 confusion matrix permutations, we show how misleading F1 is because it is not chance-adjusted.  We demonstrate this with respect to simple Bookmaker's odds.
\item The F1 metric is still widely used by researchers investigating classifiers for software defect prediction.  Our analysis of the literature suggests of the order of 800 software defect prediction papers have used this metric in the past five years alone.
\item We find that more than a fifth (21.95\%) of all results change not only in magnitude but most importantly, in conclusion (or direction) when the unbiased MCC is used, instead of the F1 metric.
\item In passing, we also note that some classifiers do not perform well, i.e., less well than chance.  This is not apparent if researchers rely on F1, although clear from a negatively valued correlation coefficient.
\item Unsurprisingly the smaller the effect (difference in performance between pairs of classifier) the more likely the conclusion will change.  The odds-ratio between the lowest and top tertile is 4.11 (95\% CI = $[ 3.65, 4.64]$).
\end{enumerate}

This tendency to use F1 in software defect studies has wider ramifications than just single studies, since it then propagates through into meta-analyses which are often based on this metric \cite{Hoss17,Malh15}.  Other meta-analyses have been obliged to discard significant amounts of data when researchers only reported results in terms of F1 or failed to provide confusion matrices \cite{Shep14,Li20}.

Finally, we wish to be clear that we are not making a criticism of the authors of the 38 primary studies included in our meta-analysis.  To their considerable credit they have provided the necessary data to make our investigation possible.  Nor are we claiming that they have relied upon the F1 metrics.  They have, however, provided the means whereby we can answer the question: how much does using F1 for software defect prediction studies matter?  Sadly, the answer seems to be: a good deal.

\subsection{Threats to validity} \label{SubSec:Threats}

\emph{Internal:} threats relate to extent to which the design of our investigation enables us to argue there is evidence to support our claim (i.e., that using F1 as a response variable for defect prediction studies causes misleading results and therefore conclusions).

\begin{enumerate}
\item Can we be sure that discordance is an appropriate way to reveal problems with study conclusions due to using F1?  We believe using the idea of sign or direction change is the most fundamental way of considering pairs of results.  In other words in terms of preference relations.  Changing a preference for Classifier A to Classifier B inverts the meaning of the results.  The alternative of setting some minimum effect magnitude threshold by comparison seems arbitrary.
\item Perhaps researchers only focus on very large differences between F1 measures?  So a change in direction or discordant results might not matter if both sets of results are very close to a zero effect size.  Researchers have indeed used a range of decision procedures to determine whether the magnitude of the effect matters (see Table \ref{Tab:Proc}) but the majority (24/38) use NHST.  One of the many criticisms of NHST is that very small effects can be `significant' when $n$ is large which is typically the case for software defect studies.  
\item Measurement error, for instance with regard to the boundary conditions e.g., F1=0.  Elsewhere we have found that reporting and/or measurement errors can be depressingly prevalent \cite{Shep19}.  However, it is not obvious why F1 would be more impacted than MCC.
\item The data sets used by researchers generally assume simple relationships and traceability between defects and repairs.  Herbold~\cite{Herb19} has argued, with some justification, that we should expect m:n relationships between defects and software units.  Whilst this may well weaken the practical relevance of software defect prediction research, our focus is on how we assess classification performance and whether it matters if F1 is employed.  Hence we believe that this, valid threat, is somewhat peripheral.
\item In line with all the research included in this study, we ignore costs by assuming the costs of false positives and false negatives are equal.  Penalising one class of error more than the other is a potentially important area of software defect prediction research \cite{Khos98,Herb19}, but one outside the scope of this study.
\end{enumerate}

\noindent
\emph{External:} threats concern the generalisability of our findings.

\begin{enumerate}
\item Is our sample large enough?  We have more than 12,000 pairs of results plus we have sought to locate all studies that provide the data we need for our investigation. 
\item Suppose we had looked at other studies? Requiring the study to publish both F1 and MCC results might skew the findings?  This is possible and it does seem that using MCC is a relatively recent practice.  We excluded the grey literature (i.e., unrefereed studies).  Given that (hopefully) research methods and practice improve over time and our focusing on demonstrably refereed studies, this would suggest that if anything we are biased to higher quality studies.  Consequently, the overall picture could conceivably be worse than our meta-analysis reveals.

\end{enumerate}

\subsection{Recommendations} \label{SubSec:Recc}

These results have implications both for researchers, but also for consumers of their research (both other researchers and practitioners).

\begin{itemize}
\item The first, and most obvious, implication is that researchers should stop using the F1 metric to analyse and compare software defect classifiers.  We should \emph{not} reason that because people have previously used F1 we should continue to do so.  Otherwise our research will be perpetually mired in the past!  Minimally, we suggest researchers should provide other unbiased metrics such as the Matthews correlation coefficient (MCC) \cite{Bald00}.  In this study we have strongly advocated for MCC. This is for three reasons, (i) it is chance corrected, (ii) like any correlation coefficient its meaning is easy to interpret and serves as a standardised effect size and (iii) it approximates a chi-square distribution.  A possible alternative is Youden's J \cite{Youd50} which indicates the performance relative to guessing or chance but it does not offer all the advantages of a correlation coefficient.  We do not, however, see AUC as a good alternative for our purposes of comparing specific classifiers since it is based on the performance of a whole family classifiers based on varying the acceptance threshold  (see Section~\ref{Subsec:PerfMetrics}). 
\item We need full reporting of data, results and code / scripts to assist with auditability \cite{Muna17,Fern19}.  Preferably, papers should provide all the confusion matrices so that a wide range of metrics can potentially be computed as secondary analysis, in the event that researchers have a preference for a particular, but alternative, metric.
\item When undertaking meta-analyses these should \emph{not} be based upon F1 results.  Instead, it may be possible to either use results based upon other metrics or derive them from other information reported \cite{Bowe14}.  Otherwise, the risk that ~22\% of the results used in a meta-analysis are completely misleading must be viewed as rendering the results critically contaminated.
\item When reading past studies based upon F1, consider the absolute size of the effect plus the confidence interval but ignore statistical significance.  Unless the absolute F1 difference is non-trivial (our analysis would suggest $\approx 0.1$) we recommend little credence should be given to such a result.  Even then, this still implies a $\sim 12\%$ chance that not only is the magnitude of the effect wrong but it is actually in the opposite direction.  So instead of classifier A being considerably better than B, it turns out that B is better than A.  How can we expect practitioners to deploy substantial resources in the real-world e.g., guiding their testing effort when such advice could be completely misguided.
\end{itemize}

It seems a great deal of research effort has been deployed on the clearly important problem of how to predict where testing effort should be focused in large software systems.  As important as that question might be, we can hardly expect our research to have much practical impact unless we, as a community, take reasonable steps to ensure our computational experiments have meaning.

\begin{appendices}
\section{Details of the primary studies included in the systematic review} \label{App:Details}

\begin{supertabular}{c p{2cm} c c p{6cm} r}
\hline
Paper & Authors & Year & Type & Title & \#Results \\ 
  \hline
\cite{Abae18} & Abaei et al.  & 2018 & J & A fuzzy logic expert system to predict module fault proneness using unlabeled data &  14 \\ 
  \cite{AlDa18} & Al Dallal & 2018 & J & Predicting fault-proneness of reused object-oriented classes in software post-releases &  27 \\
  \cite{Ali20} & Ali et al.  & 2020 & J & Software Defect Prediction Using Variant based Ensemble Learning and Feature Selection Techniques &  220 \\   
  \cite{Amas18} & Amasaki & 2018 & C & Cross-version defect prediction using cross-project defect prediction approaches: does it work? & 1404 \\ 
  \cite{Amas20} & Amasaki & 2020 & J & Cross-version defect prediction: use historical data, cross-project data, or the both? & 171 \\ 
  \cite{Anta20} & Antal et al.  & 2020 & J & Enhanced Bug Prediction in JavaScript Programs with Hybrid Call-Graph Based Invocation Metrics &  36 \\
  \cite{Ayon19} & Ayon & 2019 & C & Neural network based software defect prediction using genetic algorithm and particle swarm optimization &  30 \\ 
  \cite{Bangash20} & Bangash et al.  & 2020 & J & On the time-based conclusion stability of cross-project defect prediction models &  190 \\
  \cite{Bowe12} & Bowes et al. & 2012 & C & Comparing the performance of fault prediction models which report multiple performance measures: recomputing the confusion matrix &   6 \\ 
  \cite{Bowe15} & Bowes et al. & 2015 & C & Different classifiers find different defects although with different level of consistency &  12 \\ 
  \cite{Bowe18} & Bowes et al.  & 2018 & J & Software defect prediction: do different classifiers find the same defects? &  18 \\ 
  \cite{Chen20} & Chen et al. & 2020 & J & An empirical study on heterogeneous defect prediction approaches & 336 \\ 
  \cite{Felix20} & Felix et al. & 2020 & J & Predicting the number of defects in a new software version &  30 \\
  \cite{Ge18} & Ge et al.  & 2018 & C & Comparative study on defect prediction algorithms of supervised learning software based on imbalanced classification data sets &  84 \\ 
  \cite{Gong19} & Gong et al.  & 2019 & J & An improved transfer adaptive boosting approach for mixed?project defect prediction & 1512 \\ 
  \cite{Herb18} & Herbold et al.  & 2018 & J & A comparative study to benchmark cross-project defect prediction approaches & 1890 \\ 
  \cite{Iqba20} & Iqbal & 2020 & J & A classification framework for software defect prediction using multi-filter feature selection technique and MLP & 588 \\ 
  \cite{Lena20} & Lenarduzzi et al. & 2020 & C & Are SonarQube rules inducing bugs? &  36 \\ 
  \cite{Matl19} & Matloob et al.  & 2019 & J & A framework for software defect prediction using feature selection and ensemble learning techniques &  85 \\ 
  techniques &  85 \\ 
  \cite{Naseem20} & Naseem et al.   & 2020 & J & Investigating Tree Family Machine Learning Techniques for a Predictive System to Unveil Software Defects & 450 \\ 
  \cite{Nezh20} & Nezhad et al.   & 2020 & J & Software defect prediction using over-sampling and feature extraction based on mahalanobis distance & 180 \\ 
  \cite{Niu20} & Niu et al.  & 2020 & J & Cost-sensitive Dictionary Learning for Software Defect Prediction &  1595 \\
  \cite{Pan19} & Pan et al.  & 2019 & J & An improved cnn model for within-project software defect prediction & 336 \\ 
  \cite{Pand20} & Pandey et al. & 2020 & J & Bpdet: an effective software bug prediction model using deep representation and ensemble learning techniques  & 429 \\ 
  \cite{Pecorelli20} & Pecorelli et al.  & 2020 & J & A large empirical assessment of the role of data balancing in machine-learning-based code smell detection &  11 \\
  \cite{Rizw17} & Rizwan et al. & 2017 & C & Empirical study on software bug prediction &  78 \\ 
  \cite{Rodr14} & Rodrigues et al. & 2014 & C & Preliminary comparison of techniques for dealing with imbalance in software defect prediction & 726 \\ 
  \cite{Ship18} & Shippey et al. & 2018 & C & Code cleaning for software defect prediction: a cautionary tale &  18 \\ 
  \cite{Tian20} & Tian et al.  & 2020 & C & How Well Just-In-Time Defect Prediction Techniques Enhance Software Reliability? &  110 \\
  \cite{Tong18} & Tong et al.  & 2018 & J & Software defect prediction using stacked denoising autoencoders and two-stage ensemble learning & 219 \\ 
  \cite{Tong20} & Tong et al.  & 2020 & J & Credibility Based Imbalance Boosting Method for Software Defect Proneness Prediction &  231 \\
  \cite{Tran19} & Tran et al. & 2019 & C & Combining feature selection, feature learning and ensemble learning for software fault prediction &  36 \\ 
  \cite{Xuan15} & Xuan et al. & 2015 & C & Evaluating defect prediction approaches using a massive set of metrics: an empirical study &  15 \\ 
  \cite{Xu19} & Xu et al.  & 2019 & J & Software defect prediction based on kernel pca and weighted extreme learning machine & 237 \\ 
  \cite{Xu20} & Xu et al.  & 2020 & J & Imbalanced metric learning for crashing fault residence prediction &  237 \\
  \cite{Zhan16} & Zhang et al. & 2016 & J & Towards building a universal defect prediction model with rank transformed predictors &  50 \\ 
  \cite{Zhang20} & Zhang et al.  & 2020 & J & Automated defect identification via path analysis-based features with transfer learning &  24 \\
  \cite{Zhao19} & Zhao et al.  & 2019 & J & Siamese dense neural network for software defect prediction with small data & 900 \\ 
   \hline
\end{supertabular}

\end{appendices}

\section*{Acknowledgements}
The authors wish to thank the reviewers and the editor for their helpful and constructive comments.  They also thank the authors of the 38 primary studies included for providing sufficient information to make this analysis possible.  We also wish to stress that our criticism of F1 does not mean we are criticising their papers.  On the contrary, their foresight that alternative metrics to F1 are needed, has been invaluable.  Jingxiu Yao wishes to acknowledge the support of the China Scholarship Council.

\section*{Conflict of interest}
The authors declare that they have no conflict of interest.

\bibliography{JingXiu_IST}

\end{document}